\documentclass[pdftex,twocolumn,epjc3]{svjour3}          

\RequirePackage[T1]{fontenc}

\smartqed  

\usepackage[T1]{fontenc}
\usepackage{lmodern}
\usepackage{amsmath}
\usepackage{amsfonts}
\usepackage{amssymb}
\usepackage{mathrsfs}
\usepackage{physics}
\usepackage{graphicx}
\usepackage{caption}
\usepackage{subcaption}
\usepackage{enumitem}
\usepackage{framed,float}
\usepackage{booktabs}
\usepackage{array}
\usepackage{soul}
\usepackage{multirow}
\usepackage{comment}
\usepackage{braket}
\usepackage{todonotes}
\usepackage{youngtab}
\usepackage{xcolor}
\usepackage{tikz}
\usepackage{color}
\usepackage{extarrows}
\usetikzlibrary{calc,arrows,decorations.markings}
\usepackage{CJKutf8}
\usepackage{hyperref}

\renewcommand{\makeheadbox}{}

\def \be  {\begin{equation}}
\def \ee  {\end{equation}}
\def \ba {\begin{equation}\begin{aligned}}
\def \ea {\end{aligned}\end{equation}}
\def \bea  {\begin{eqnarray}}
\def \eea  {\end{eqnarray}}

\begin{document}


\title{\boldmath Thermal Breaking of the I-Love Universality for Hot White Dwarfs}

\author{Jinyi Lv\thanksref{e1,addr1,addr2}
\and
Jing-Yi Wu\thanksref{e2,addr3}
	\and
    Hongji Chen\thanksref{e3,addr2}
	\and
    Kexin Jia\thanksref{e4,addr2}
	\and
    Weihan Sun\thanksref{e5,addr2}
	\and
	Kilar Zhang\thanksref{e6,addr2,addr4,addr5}
}

\thankstext{e1}{e-mail: lvjinyi@shu.edu.cn}
\thankstext{e2}{e-mail: wujingyi222@mails.ucas.ac.cn (corresponding author)}
\thankstext{e3}{e-mail: 
jkx0116@163.com }
\thankstext{e4}{e-mail: admissions@shu.edu.cn }
\thankstext{e5}{e-mail: swh114514@163.com }
\thankstext{e6}{e-mail: kilar@shu.edu.cn (corresponding author)}

\institute{
    State Key Laboratory of Precision Spectroscopy, Joint Institute of Advanced Science and Technology, School of Physics and Electronic Science, East China Normal University, Shanghai 200062, China\label{addr1}
    \and
	Department of Physics and Institute for Quantum Science and Technology, Shanghai University, Shanghai 200444, China\label{addr2}
	\and
	School of Astronomy and Space Science, University of Chinese Academy of Sciences (UCAS), Beijing 100049, China\label{addr3}
	\and
	Shanghai Key Lab for Astrophysics, Shanghai 200234, China\label{addr4}
	\and
	Shanghai Key Laboratory of High Temperature Superconductors, Shanghai 200444, China\label{addr5}
}

\date{}

\maketitle

\begin{abstract}

The universal I-Love-Q relations for compact stars have significant applications in gravitational-wave astronomy, but thermal effects can break these relations in low-mass white dwarfs. In this work, we employ the stellar evolution code MESA to construct realistic models of $0.15 \, M_{\odot}$ helium-core and $0.6 \, M_{\odot}$ carbon-oxygen core white dwarfs at various temperatures. By utilizing the Clairaut-Radau equation, we quantitatively extract the radial variation of the eccentricity of internal isodensity surfaces. Our numerical results demonstrate that higher central temperatures amplify the eccentricity variation, causing the I-Love relations to deviate from the zero-temperature Chandrasekhar model, whereas subsequent cooling restores them. This confirms that the temperature-induced violation of the universal relations is fundamentally driven by the loss of self-similarity in isodensity surfaces, providing key insights into the applicability conditions of I-Love-Q relations in compact objects.
\end{abstract}

\section{Introduction}
\label{sec:intro}

The universal $I$-$\text{Love}$-$Q$ relations, which connect the dimensionless moment of inertia ($\bar{I}$), the tidal Love number ($\bar{\lambda}$), and the quadrupole moment ($\bar{Q}$), represent a cornerstone in the study of compact objects~\cite{yagi2013love,yagi2013love_science}. These relations are remarkably insensitive to the internal equation of state (EOS) for cold neutron stars and isolated white dwarfs, offering a powerful tool for testing general relativity and breaking degeneracies in gravitational-wave data analysis.

For typical cold, carbon-oxygen white dwarfs, the Chandrasekhar zero-temperature EOS yields a robust $I$-$\text{Love}$-$Q$ relation. However, recent developments in astronomical observations have turned the spotlight onto low-mass helium-core white dwarfs and hot, newly born white dwarfs, where thermal effects play a non-negligible role. Boshkayev et al.~\cite{boshkayev2016love} first noted that the inclusion of finite-temperature effects induces noticeable deviations from the standard cold universal relations. Subsequently, Taylor, Yagi, and Arras~\cite{taylor2020love} carried out pioneering work by computing the $I$-$\text{Love}$-$Q$ relations for realistic white dwarfs using the Modules for Experiments in Stellar Astrophysics (MESA) code. 

Despite these advancements, a key piece of the puzzle remains missing. While Taylor et al.~\cite{taylor2020love} demonstrated that realistic EOS variations affect the relations, a systematic, quantitative investigation into how the thermal evolution of a white dwarf explicitly drives the breaking of this universality remains lacking. More importantly, the underlying physical mechanism has not been thoroughly understood from a structural geometry perspective. Yagi et al.~\cite{yagi2014love,stein2014three} hypothesized that the mathematical origin of the $I$-$\text{Love}$-$Q$ universality lies in the "approximate self-similarity of isodensity surfaces" inside the star. When a white dwarf is hot, the degeneracy of electrons is partially lifted, altering the radial pressure profile and potentially destroying this geometric self-similarity. A rigorous connection between temperature, the breakdown of isodensity self-similarity, and the resultant residuals in the universal relations has yet to be established for realistic white dwarf models.

To address these open questions, this work provides a comprehensive analysis of the thermal breaking of $I$-$\text{Love}$ universality across different evolutionary stages of white dwarfs. Utilizing the state-of-the-art stellar evolution code MESA, we construct realistic tracks for both $0.15 \, M_{\odot}$ helium-core and $0.6 \, M_{\odot}$ carbon-oxygen core white dwarfs spanning a wide range of central temperatures. By solving the perturbed stellar structure equations alongside the Clairaut-Radau equation, we quantitatively extract the radial variation of the eccentricity of internal isodensity surfaces ($e_s/e_c$). 

Our results clearly demonstrate that higher central temperatures amplify the radial variation of eccentricity, thereby directly breaking the geometric self-similarity of the internal layers. This structural deformation serves as the physical bridge explaining why the $I$-$\text{Love}$ relations deviate significantly from the zero-temperature Chandrasekhar model. Although we focus primarily on the tracking of the $I$-$\text{Love}$ relation due to its direct relevance to tidal deformability, the geometric insight gained here applies universally to the broader $I$-$\text{Love}$-$Q$ framework.

The structure of this paper is as follows: Section.~\ref{sec:static_sol} briefly reviews the static solutions of stars and two types of EOS; Section.~\ref{sec:ilq_calc} introduces the calculation methods for the moment of inertia, tidal Love number, and rotating quadrupole moment; Section.~\ref{sec:struct_isodensity} summarizes the structural origin of the I-Love-Q relations in neutron stars, with emphasis on isodensity-surface self-similarity and have a simple EOS test before the MESA models; Section.~\ref{sec:hdwarf} presents the MESA models for helium-core and carbon-oxygen-core white dwarfs, the eccentricity extraction algorithm, and the numerical results of the I-Love relations, followed by analysis and discussion; finally, the Conclusions and Outlook section summarizes the paper and discusses future work.

\section{Static Solution of Stars}
\label{sec:static_sol}
The static solution of a star provides the structural information of the star, which serves as the necessary background for further solving physical quantities such as eccentricity, moment of inertia, tidal Love number, and rotating quadrupole moment. 

\subsection{TOV Equations and the Newtonian Approximation}
 The purpose of this section is to fix the notation and the normalizations used below, while keeping only the formulae needed for the later white-dwarf calculation.  We use geometrized units $G=c=1$ unless otherwise specified.  The stellar radius and mass are denoted by $R_*$ and $M_*$, and the compactness is
\begin{equation}
   \hfill C \equiv \frac{M_*}{R_*} . \hfill
\end{equation}

The reference stellar model considered in this work is a slowly rotating and weakly tidally deformed relativistic star. The deformation is assumed to be sufficiently small so
that all quantities can be expanded around a static
Tolman-Oppenheimer-Volkoff equations (TOV), where \(p\) is the pressure and \(\rho\) is the energy density  \cite{ott2013static},

\begin{equation}
  \hfill \frac{dm}{dr}=4\pi r^{2}\rho, \hfill
\end{equation}
\begin{equation}
  \hfill \frac{dp}{dr}=-\frac{(\rho+p)\left[m+4\pi r^{3}p\right]}{r(r-2m)}, \hfill 
\end{equation}
\begin{equation}
  \hfill \frac{d\nu}{dr}=\frac{2\left[m+4\pi r^{3}p\right]}{r(r-2m)}. \hfill
\end{equation}

Although the general framework allows for a relativistic star, for the hot white dwarfs considered in this work, the compactness is extremely small,  thus the Newtonian hydrostatic equations are sufficient. For low-mass white dwarfs, we can simplify the TOV equations and use the hydrostatic equilibrium equation under the Newtonian approximation \cite{boshkayev2016love}:
\begin{equation}
    \hfill \frac{dp}{dr} = -\rho \frac{m}{r^2}, \hfill
\end{equation}
\begin{equation}
    \hfill \frac{dm}{dr} = 4\pi r^2 \rho. \hfill
\end{equation}

\subsection{Stellar Equations of State}
When numerically solving the internal structure of a star, the hydrostatic equilibrium equations (such as the hydrostatic equation in the Newtonian framework or the TOV equations in general relativity) alone are insufficient to close the system of differential equations. We also need to introduce an EOS that describes the properties of the matter inside the star, i.e., the relationship between pressure \(p\) and mass density (or energy density) \(\rho\): \(p = p(\rho)\). Only by combining the hydrostatic equilibrium equations with the EOS can we integrate outward from the center and completely determine the radial distributions of density, pressure, mass, etc., thereby obtaining the macroscopic structure of the star (such as the mass-radius relation). We consider the following two EOS.

\subsubsection{Polytropic Equation of State}

The polytropic EOS \cite{yagi2013love,yagi2013love_science}, is a simplified relationship obtained through mathematical fitting. Its general form is:
\begin{equation}
\hfill p = K \rho^{(1 + 1/n)}\equiv K \rho^{\gamma},  \hfill
\label{eq:poly}
\end{equation}
where \(K\) is a constant, and \(n\) is the polytropic index. It ignores complex factors such as temperature and composition, summarizing the microscopic thermodynamic properties into a simple power-law relationship. Therefore, it is widely used in analytical and semi-analytical studies to approximate the structure of realistic stars (e.g., neutron stars and white dwarfs).

\subsubsection{Chandrasekhar Zero-Temperature Model}

The Chandrasekhar EOS is the simplest EOS for describing the structure of white dwarfs \cite{chandrasekhar1931maximum,chandrasekhar1989selected}. Under zero-temperature conditions, this equation is constructed based on a completely degenerate electron Fermi gas, ignoring the contribution of ionic pressure. The total pressure is dominated by the electron pressure, while the energy density mainly comes from the rest mass energy of atomic nuclei \cite{boshkayev2016love}.

The relationship between the electron number density \(n_e\) and the Fermi momentum \(P_e^F\) is:
\begin{equation}
\hfill n_e = \frac{(P_e^F)^3}{3\pi^2 \hbar^3} = \frac{(m_e c)^3}{3\pi^2 \hbar^3} x_e^3, \hfill
\end{equation}
where \(x_e = \frac{P_e^F}{m_e c}\) is the dimensionless Fermi momentum, \(m_e\) is the electron mass, and \(c\) is the speed of light. From this, the electron pressure is:
\begin{equation}
\begin{aligned}
     \quad \quad \quad \quad P_e &= \frac{m_e^4 c^5}{8\pi^2 \hbar^3} \Big( x_e \sqrt{1+x_e^2} \\ 
     &\left(\frac{2}{3}x_e^2 - 1\right) + \ln\left(x_e + \sqrt{1+x_e^2}\right) \Big), 
\end{aligned}
\end{equation}

This expression reduces to \(P_e \propto n_e^{5/3}\) in the non-relativistic limit \((x_e \ll 1)\) and asymptotically approaches \(P_e \propto n_e^{4/3}\) in the ultra-relativistic limit \((x_e \gg 1)\), which is the origin of the Chandrasekhar mass limit.

Thus, using \(x_e\) as a free parameter, \(P_e\) is determined by the above equation, and \(\rho\) is obtained from \(n_e\), establishing a parameterized relationship between pressure and density. This zero-temperature model is suitable for cold white dwarfs, i.e., those whose internal temperature is much lower than the Fermi temperature, and it is also the mathematical foundation of the classical Chandrasekhar theory of white dwarfs.

\section{Theoretical Framework for I-Love-Q Perturbations}
\label{sec:ilq_calc}

 In this section we only collect the perturbative equations and normalizations needed below. The background quantities $m(r)$, $p(r)$, and $\rho(r)$ are those obtained from the static solutions in Section.~\ref{sec:static_sol}.

\subsection{The Moment of Inertia}
For uniform slow rotation with angular velocity $\Omega$, the linear-order perturbation is described by the parameter $\omega(r)$. It satisfies \cite{yagi2013love}
\begin{equation}
\frac{d^{2}\omega}{dr^{2}} + \frac{4\left[1 - \pi r^{2}(\rho+p)e^{\lambda}\right]}{r}\frac{d\omega}{dr} - 16\pi(\rho+p)e^{\lambda}\omega = 0,
\label{eq:frame_dragging}
\end{equation}
where $\phi(r)=-\ln(1-2m/r)$ is the metric function. In the exterior region,
\begin{equation}
\hfill \omega(R_*) = \Omega - \frac{2S}{R_*^{3}} = \Omega\left(1 - \frac{2I}{R_*^{3}}\right), \hfill
\end{equation}
where $S$ is the angular momentum and $I\equiv S/\Omega$. Matching the regular interior solution to this exterior form gives
\begin{equation}
\hfill I = \frac{8\pi}{3\Omega}\int_{0}^{R_*}\frac{e^{-(\nu+\phi)/2}r^{5}(\rho+p)\omega}{r-2m}\,dr. \hfill
\label{eq:I_rel}
\end{equation}
For white dwarfs, due to their low compactness, general relativistic effects are negligible; therefore, we compute the moment of inertia using the Newtonian approximation \cite{hartle1967slowly,hartle1968slowly}. To facilitate calculations from the mass grid output by MESA, we replace the integration variable with the mass element instead of the radius element, thereby rewriting it in the following form \cite{yagi2013love}:
\begin{equation}
    \hfill I^{\mathrm{N}} = \frac{2}{3}\int_{0}^{M} r^{2} dm. \hfill
\end{equation}
The $dm$ in the equation can be obtained by differentiating the mass column in the MESA output data.
The dimensionless moment of inertia used in the I-Love-Q relations is
\begin{equation}
\hfill \overline{I} \equiv \frac{I}{M_*^3}. \hfill
\end{equation}

\subsection{The Rotational Quadrupole Moment}
At order $\mathcal{O}(\Omega^2)$, rotation induces an oblate quadrupolar deformation. In the Hartle-Thorne expansion this deformation is represented by the $l=2$ even-parity perturbations, and matching the interior and exterior solutions fixes the integration constant $A$ \cite{hartle1967slowly,hartle1968slowly}. The spin-induced quadrupole moment is then \cite{yagi2013love}
\begin{equation}
\hfill Q = -\frac{S^{2}}{M_*} - \frac{8}{5}AM_*^{3}. \hfill
\end{equation}
The corresponding spin-normalized quantity is
\begin{equation}
\hfill \bar{Q} \equiv -\frac{QM_*}{S^{2}} = 1 + \frac{8}{5}AM_*^{4}, \hfill
\end{equation}
where the sign convention makes $\bar{Q}$ positive for an oblate configuration.

\subsection{Tidal Deformability}
For a star in an external quadrupolar tidal field $\mathcal{E}_{ij}$, the induced quadrupole moment satisfies
\begin{equation}
\hfill Q_{ij}=-\lambda\mathcal{E}_{ij}. \hfill
\end{equation}
The static $l=2$ polar perturbation can be reduced to a second-order equation for $H_2(r)$ in the metric perturbation, defined in \cite{hinderer2008tidal}:
\begin{equation}
\begin{aligned}
    &\frac{d^{2}H_{2}}{dr^{2}} + \left[\frac{2}{r} + \left(\frac{2m}{r^{2}} + 4\pi r(p-\rho)\right)e^{\phi}\right]\frac{dH_{2}}{dr} - \\
    &\left[\frac{6e^{\phi}}{r^{2}} - 4\pi e^{\phi}\left(5\rho + 9p + (\rho+p)\frac{d\rho}{dp}\right) + \left(\frac{d\nu}{dr}\right)^{2}\right]H_{2} = 0.
\end{aligned}
\label{eq:H2_diff}
\end{equation}
Introducing $y(r)\equiv rH_2'(r)/H_2(r)$ gives the first-order form \cite{Postnikov_2010}
\begin{equation}
\begin{aligned}
    &\frac{dy}{dr} + \frac{y(y+1)}{r} + ye^{\phi}\left(\frac{2m}{r^{2}} + 4\pi r(p-\rho)\right) 
- \\
&\frac{6e^{\phi}}{r} + 4\pi e^{\phi}\left(5\rho + 9p + (\rho+p)\frac{d\rho}{dp}\right) + r\left(\frac{d\nu}{dr}\right)^{2} = 0,
\end{aligned}
\label{eq:tidal-y-equation}
\end{equation}
with regular central boundary condition $y(0)=2$. The surface value is denoted by $y_R\equiv y(R_*)$.

The dimensionless tidal deformability is
\begin{equation}
\hfill \overline{\lambda} \equiv \frac{\lambda}{M_*^5} = \frac{2}{3}k_{2}C^{-5}, \hfill
\label{eq:lambda_dimless}
\end{equation}
where the relativistic quadrupolar Love number $k_2$ is \cite{hinderer2008tidal}

\begin{equation}
    \begin{aligned}
        k_{2} &= \frac{8C^{5}}{5}(1-2C)^{2}\left[2 + 2C(y_{R}-1) - y_{R}\right]\mathcal{D}^{-1}, \\
        \mathcal{D} &= 2C\left[6 - 3y_{R} + 3C(5y_{R}-8)\right] + \\
        & 4C^{3}\left[13 - 11y_{R} + C(3y_{R}-2) + 2C^{2}(1+y_{R})\right] \\ 
        & \quad + 3(1-2C)^{2}\left[2 - y_{R} + 2C(y_{R}-1)\right]\ln(1-2C).
\end{aligned}
\label{eq:k2-gr}
\end{equation}

At low compactness, expanding Eq.~\eqref{eq:k2-gr} in $C$ to the leading order yields the Newtonian result:

\begin{equation}
	\hfill k_2 = \frac{1}{2} \frac{2 - y_R}{3 + y_R},\hfill
\end{equation}
and the dimensionless tidal Love number is then given by Eq.~(\ref{eq:lambda_dimless}).

\section{Structural Foundations and Isodensity Self-Similarity of I-Love-Q Relations}
\label{sec:struct_isodensity}

In this section, we summarize the core concepts and arguments introduced in \cite{yagi2014love}, which serve as a foundation to motivate our discussion in Section~\ref{sec:hdwarf} regarding the conditions under which the I-Love relations break down in hot white dwarfs. Furthermore, prior to utilizing the MESA code, we perform a simplified analysis to provide preliminary validation for this behavior.

\subsection{Regions Contributing Most to \(I\) and \(Q\) in Neutron Stars}

\cite{yagi2014love} employs a piecewise polytropic EOS (i.e., splicing together segments of the form Eq.~\eqref{eq:poly}). By varying the parameters of each segment, it is found that the regions contributing most to the moment of inertia \(I\) and the quadrupole moment \(Q\) in neutron stars lie between \(50\%\) and \(95\%\) of the stellar radius, corresponding to a density range of \(10^{14} \sim 10^{15}~\mathrm{g/cm^{3}}\). Furthermore, it points out that although current neutron-star EOS (e.g., SLy4, APR) differ by as much as \(17\%\) in this region, the deviation in the I-Love-Q relations remains at the level of \(\sim1\%\). 

\subsection{Isodensity-surface Self-similarity}

The most important conclusion in \cite{yagi2014love} is that the self-similarity of isodensity surfaces is the direct cause of the I-Love-Q relations. Indeed, it is proved in \cite{stein2014three} that under the condition of strictly self-similar isodensity ellipsoids, universal relations generally exist among the multipole moments of a star. \cite{yagi2014love} further decomposes this hypothesis into three sub-conditions: isodensity-surface self-similarity, ellipsoidal shape, and a spherically symmetric density profile. It is argued that the most critical of these is the self-similarity of isodensity surfaces ; once this condition is relaxed, the I-Love-Q universal relations are broken.

In addition, by analyzing the eccentricity distribution inside slowly and rapidly rotating neutron stars and quark stars, \cite{yagi2014love} finds that within the dominant contribution region, the eccentricity varies by only \(10\%\)–\(30\%\), and the deviation of the I-Love-Q relations between rapidly and slowly rotating stars is only of order \(0.1\%\). In contrast, for non-compact stars (e.g., ordinary stars), the eccentricity of isodensity surfaces varies by more than \(100\%\), and in that case the I-Love-Q relations completely fail.

In \cite{yagi2014love}, two extreme cases are studied: small eccentricity variation (neutron stars, quark stars) and huge eccentricity variation (ordinary stars); While for white dwarfs, the eccentricity variation lies between these two extremes. This is one of the motivations for our study on how the eccentricity variation in white dwarfs affects the I-Love-Q relations.

\subsection{A Controlled Test of the Eccentricity-Variation Criterion in White Dwarfs}

\subsubsection{A Simple EoS Test before the MESA Models}

Before studying realistic MESA white-dwarf models, we first perform a simple
controlled test.  The goal is to isolate one structural question: if the
radial stratification of a white-dwarf EOS is changed by
a single parameter, does the I--Love residual follow the radial variation of
the eccentricity of isodensity surfaces? 

We use the composite-polytrope white-dwarf EOS discussed by Van Gorder and
Fisher~\cite{van2023spatial}, following the fitting form introduced by
Paczynski~\cite{paczynski1976common}.  With the normalized density
\(q=\rho/\rho_c\), the normalized pressure is
\begin{equation}
  \hfill  P(q)=
    \left(
    \frac{1+K}{q^{-10/3}+Kq^{-8/3}}
    \right)^{1/2},
    \qquad q=\frac{\rho}{\rho_c}.
    \label{eq:eos} \hfill
\end{equation}
Equivalently,
\begin{equation}
   \hfill P(q)=
    \sqrt{1+K}\,
    \frac{q^{5/3}}{\sqrt{1+Kq^{2/3}}}.
    \label{eq:eos_rewrite} \hfill
\end{equation}
The normalization gives \(P(1)=1\), so \(P(q)\) can be regarded as
\(P/P_c\).  In the numerical calculation, the central pressure scale is fixed
by the Chandrasekhar zero-temperature pressure at the same central density.
Thus \(K\) changes the radial shape of the EOS rather than simply rescaling
the pressure.

Equation~\eqref{eq:eos_rewrite} shows why this EOS is useful.  In the
low-density outer region, \(Kq^{2/3}\ll1\), one obtains
\begin{equation}
   \hfill P(q)\simeq \sqrt{1+K}\,q^{5/3}, \hfill
\end{equation}
so that
\begin{equation}
  \hfill  P\propto \rho^{5/3}. \hfill
\end{equation}
This is the non-relativistic degenerate-electron limit, corresponding to
\(\gamma=5/3\) or polytropic index \(n=3/2\).

In the opposite limit, \(Kq^{2/3}\gg1\), one obtains
\begin{equation}
   \hfill P(q)\simeq
    \sqrt{\frac{1+K}{K}}\,q^{4/3}, \hfill
\end{equation}
so that
\begin{equation}
   \hfill P\propto \rho^{4/3}. \hfill
\end{equation}
This is the relativistic degenerate-electron limit, corresponding to
\(\gamma=4/3\) or \(n=3\).  The transition occurs roughly at
\begin{equation}
   \hfill Kq^{2/3}\sim1,
    \qquad q_{\rm trans}\sim K^{-3/2}. \hfill
\end{equation}
Therefore larger \(K\) makes a larger part of the star approach the softer
\(\gamma=4/3\) behavior.

The local effective polytropic index is
\begin{equation}
   \hfill \gamma_{\rm eff}
    =
    \frac{d\ln P}{d\ln\rho}
    =
    \frac{5}{3}
    -
    \frac{1}{3}
    \frac{Kq^{2/3}}{1+Kq^{2/3}} .
    \label{eq:gamma_eff} \hfill
\end{equation}
Thus \(\gamma_{\rm eff}\) smoothly decreases from \(5/3\) to \(4/3\).
Changing \(K\) directly changes the radial stiffness profile and therefore
the eccentricity profile.

\subsubsection{Numerical Setup}

For each \(K\), we solve the Newtonian hydrostatic equilibrium equations for
a sequence of central densities.  We then compute the dimensionless moment of
inertia \(\bar I\), tidal deformability \(\bar\lambda\), quadrupole
\(\bar Q\), and the eccentricity profile from the Clairaut--Radau equation. A more detailed discussion of the Clairaut--Radau equation is in Section \ref{sec:Extraction of the eccentricity}.
The eccentricity variation here is measured by
\begin{equation}
   \hfill \Delta e =
    1-\frac{e_{\rm in}}{e_{\rm out}} .
    \label{eq:ecc_var} \hfill
\end{equation}
A larger \(\Delta e\) means a stronger loss of self-similarity of isodensity
surfaces.

We take \(K=0.684732\) as the reference sequence because it approximates the
Chandrasekhar white-dwarf density profile well in
\cite{van2023spatial}.  The I--Love residual is defined at fixed
\(\bar\lambda\) as
\begin{equation}
   \hfill \Delta_I =
    \frac{
    \bar I-\bar I_{K=0.684732}(\bar\lambda)
    }{
    \bar I_{K=0.684732}(\bar\lambda)
    } .
    \label{eq:residual} \hfill
\end{equation}
For the fixed-mass test, we sample \(K\ge0.684732\) densely and keep one
model closest to \(0.6\,M_\odot\) for each \(K\). 

\subsubsection{Results}
\label{4.3.3}
\begin{figure*}
    \centering
    \includegraphics[width=\textwidth]{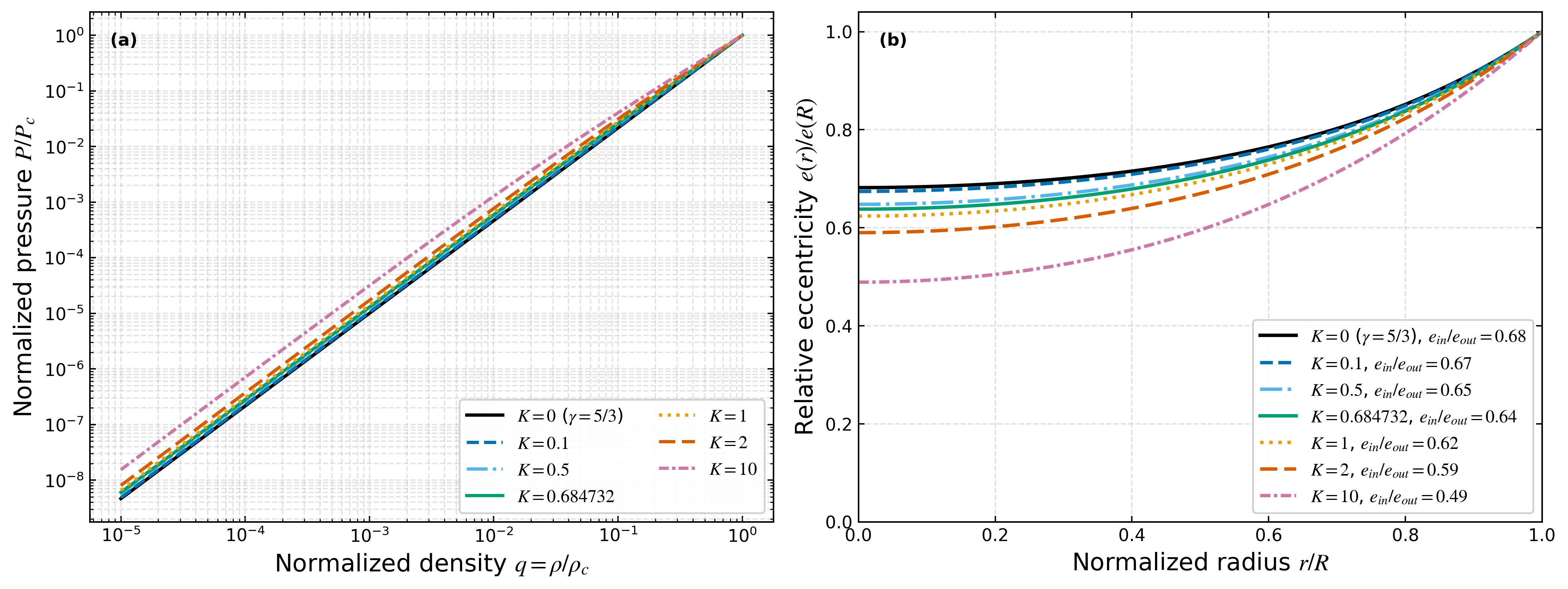}
    \caption{Composite-polytrope EOS and eccentricity profiles.  Left:
    normalized pressure \(P/P_c\) as a function of normalized density
    \(q=\rho/\rho_c\).  Right: eccentricity profiles near
    \(0.6\,M_\odot\).  Increasing \(K\) reduces
    \(e_{\rm in}/e_{\rm out}\), indicating a stronger radial variation of
    isodensity-surface eccentricity.}
    \label{fig:eos_ecc}
\end{figure*}

\begin{figure*}
    \centering
    \includegraphics[width=\textwidth]{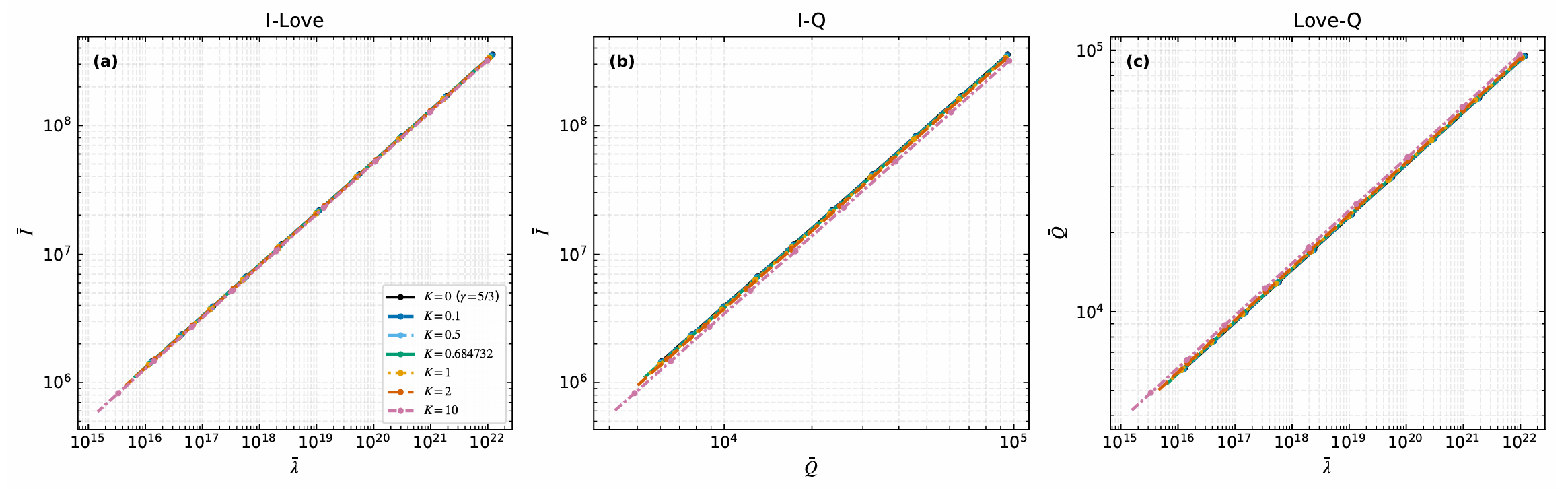}
    \caption{I--Love--Q relations for the composite-polytrope white-dwarf
    family.  The curves remain close, but their systematic separation shows
    that different \(K\) values produce measurable residuals.}
    \label{fig:iloveq}
\end{figure*}

\begin{figure}
    \centering
    \includegraphics[width=\columnwidth]{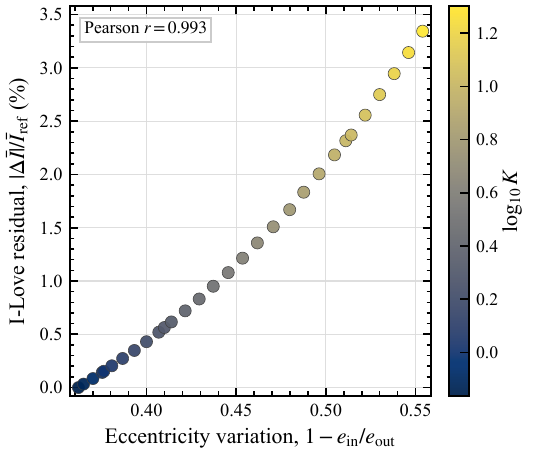}
\caption{Fixed-mass single-point sequence.  For each \(K\), only the
model closest to \(0.6\,M_\odot\) is retained.  The plotted quantity is
the absolute I--Love residual relative to the \(K=0.684732\) reference
sequence, evaluated at the same \(\bar{\lambda}\).  Consequently, the
\(K=0.684732\) reference point has zero residual by definition.  The
residual is strongly positively correlated with the eccentricity
variation \(1-e_{\rm in}/e_{\rm out}\), with Pearson coefficient
\(r=0.993\).}
    \label{fig:residual_ecc}
\end{figure}

Fig.~\ref{fig:eos_ecc} shows that increasing \(K\) changes the EOS
stratification and reduces \(e_{\rm in}/e_{\rm out}\).  This means that the
inner and outer isodensity surfaces become less self-similar.  The same
ordering appears in the I--Love--Q curves in Fig.~\ref{fig:iloveq}: the
relations are still close to universal, but the deviations are systematic.

The key result is shown in Fig.~\ref{fig:residual_ecc}.  At nearly fixed
mass, the absolute I--Love residual relative to the \(K=0.684732\)
reference sequence, evaluated at the same dimensionless tidal
deformability \(\bar{\lambda}\), is strongly positively correlated with
the eccentricity variation \(1-e_{\rm in}/e_{\rm out}\).  The reference
model has zero residual by definition.  Including the reference point,
the Pearson correlation coefficient is \(r=0.993\); excluding it gives
\(r=0.993\) to the quoted precision.

Thus, when the eccentricity changes more strongly from the stellar center to
the surface, the I--Love relation deviates more from the reference sequence.

This supports the interpretation that the I--Love--Q deviation is controlled
by the loss of self-similarity of isodensity surfaces.  In this simple test,
the microscopic thermal physics is absent; only the EOS stratification is
varied.  Therefore the result provides a useful baseline for later MESA
models.

\section{Hot White Dwarf I-Love-Q}
\label{sec:hdwarf}

K. Boshkayev et al. first demonstrated that thermal effects can break the I-Love-Q relations in white dwarfs \cite{boshkayev2018non}, with the underlying physical mechanisms to be explored. In this paper, we adopt the theory of self-similar isodensity surfaces proposed in \cite{yagi2014love} to systematically investigate how temperature alters the eccentricity distribution of internal layers. Through this approach, we elucidate the physical origin of how thermal effects disrupt the universality of the I-Love relation.

\subsection{He-core and CO-core White Dwarf MESA Models}
To construct a static white dwarf structure model, we used the open-source stellar structure and evolution code MESA \cite{taylor2020love}. By solving the one-dimensional equations of hydrostatic equilibrium, energy transport, and element evolution, MESA can self-consistently simulate the complete evolutionary path of a star from the main sequence to white dwarf cooling.

For the helium-core white dwarf model, we modified the \text{"make\_he\_wd"} test suite in the MESA package. The specific evolution steps are as follows: First, MESA evolves a pre-main-sequence star model with a mass of $1.5\,M_{\odot}$ until its internal helium core mass reaches a preset value (this paper adopts a helium-core white dwarf of $0.15\,M_{\odot}$). The helium core mass is defined as the outermost position where the hydrogen abundance drops below $1\%$. Subsequently, MESA strips off the excess mass from the outer layers of the star through a rapid mass loss process ($\Delta M = 1.35\,M_{\odot}$), leaving only the helium core. After that, the code "relaxes" the helium abundance of the entire star to $99\%$ and makes the element abundance uniformly distributed within the star, resulting in a chemically homogeneous helium-core white dwarf. Finally, the white dwarf enters the isolated cooling phase. It should be noted that in real white dwarfs, the small amount of hydrogen burning in the central region slow down the cooling process, and real white dwarfs may have thicker hydrogen envelopes than the MESA model, thus generating more heat from hydrogen burning and leading to a longer cooling time \cite{althaus1997evolution,driebe1998evolution,nelemans2000reconstructing}.

\begin{figure}[h]
    \centering
    \includegraphics[scale=0.5]{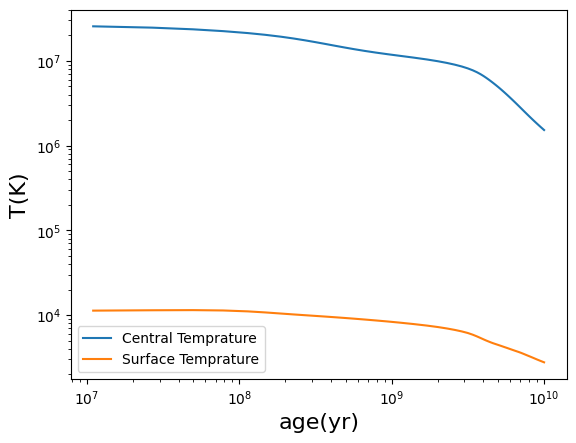}
    \caption{the cooling trace of the $0.15M_\odot$ He-core white dwarf}
    \label{1}
\end{figure}

In Fig \ref{1}, we show the evolution of the central temperature and surface temperature of this helium-core white dwarf as a function of its age. The initial central temperature of the white dwarf is approximately $10^7 \, \text{K}$, decreasing to $10^6 \, \text{K}$ after $10 \, \text{Gyr}$, while the surface temperature slowly declines from about $10^4 \, \text{K}$ to $2.7 \times 10^3 \, \text{K}$.

\begin{figure}[h]
    \centering
    \includegraphics[scale=0.5]{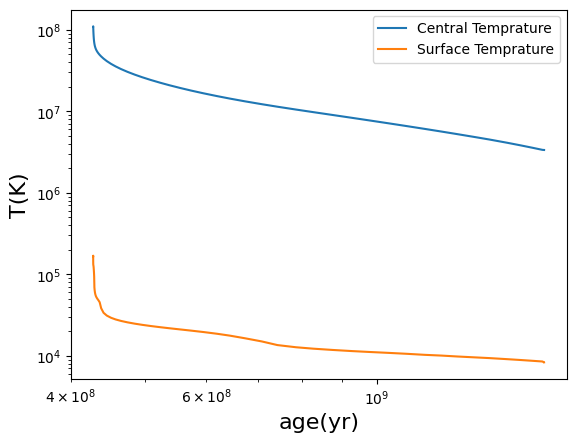}
    \caption{the cooling trace of the $0.6M_\odot$ CO-core white dwarf}
    \label{2}
\end{figure}

In addition, we also constructed a carbon-oxygen core white dwarf with a mass of about $0.6\,M_{\odot}$, which evolved from a main-sequence star of mass $3.1\,M_{\odot}$. Typically, a main-sequence star with a mass between $1.0\,M_{\odot}$ and $6.5\,M_{\odot}$ will eventually evolve into a carbon-oxygen core white dwarf \cite{camisassa2022evolution}. As shown in Fig \ref{2}, its central temperature rapidly decreases from an initial $10^8 \, \text{K}$ to $10^7 \, \text{K}$ within $10^8$ years, with a cooling rate significantly faster than that of the helium-core white dwarf \cite{mochkovitch1983freezing}.

\subsection{Extraction of the Eccentricity of White Dwarfs Isodensity Surfaces}
\label{sec:Extraction of the eccentricity}
This paper uses the Clairaut-Radau equation to extract the eccentricity of equidensity surfaces of white dwarfs. This equation is applicable to slowly and uniformly rotating stars and can be simplified to the following form \cite{tassoul2015theory}:
\begin{equation}
    \hfill a\frac{d\eta}{da} + 6\frac{\rho(a)}{\rho_m(a)}(\eta+1) + \eta(\eta-1) = 6, \hfill
\end{equation}

where
\begin{equation}
    \hfill \eta(a) = \frac{a}{e}\frac{d\epsilon}{da}, \hfill
\end{equation}

$\epsilon$ is the geometric oblateness of the isodensity surface. The average density in the equation is defined as
$$\rho_{m}(a) = \frac{3}{a^{3}} \int_{0}^{a} \rho(a') a'^{2} da',$$

Here, $\rho_m(a)$ is the average density of the matter enclosed inside a surface with mean radius $a$, and $\rho(a)$ is the mass density at the mean radius $a$. Since we assume that the star rotates slowly and uniformly, and that the deviation from spherical symmetry is very small, the difference between the mean radius $a$ and the spherical coordinate radius $r$ can be neglected. Consequently, $\rho(r)$ can be read directly from the MESA output.

For the Clairaut-Radau equation, different rotational angular frequencies of the star only scale the numerical value of the geometric oblateness of the isodensity surface proportionally, but do not change the ratio of the geometric oblateness between the outer layer and the center \cite{tassoul2015theory}. The relationship between the eccentricity ratio of the isodensity surface and the geometric oblateness ratio is as follows:

\begin{equation}
    \hfill \frac{e_s}{e_c} = \sqrt{\frac{\epsilon_s}{\epsilon_c}}, \hfill
\end{equation}

where $e_s$ represents the eccentricity of the outermost layer, and $e_c$ is the eccentricity at the center of the star.

Therefore, by taking the ratio of the eccentricity of the outer layer to that at the center, one can dimensionlessly characterize the radial variation amplitude of the eccentricity of the star's isodensity surfaces: the closer the ratio is to 1, the more similar the shapes of the isodensity surfaces (better self-similarity); the larger the ratio, the more significant the elliptical deformation of the outer layers relative to the center (breakdown of self-similarity).

\subsection{The Combined Effects of Temperature and Fermi Temperature}
To further investigate how temperature affects the eccentricity of the isodensity surfaces, we consider both the local temperature and the Fermi temperature here. From the center to the surface of a white dwarf , the temperature decreases(Figs.\ref{1} and \ref{2}); the Fermi temperature also decreases as the particle number density drops \cite{d1990cooling}. 

\begin{figure}[h]
    \centering
    \includegraphics[scale=0.3]{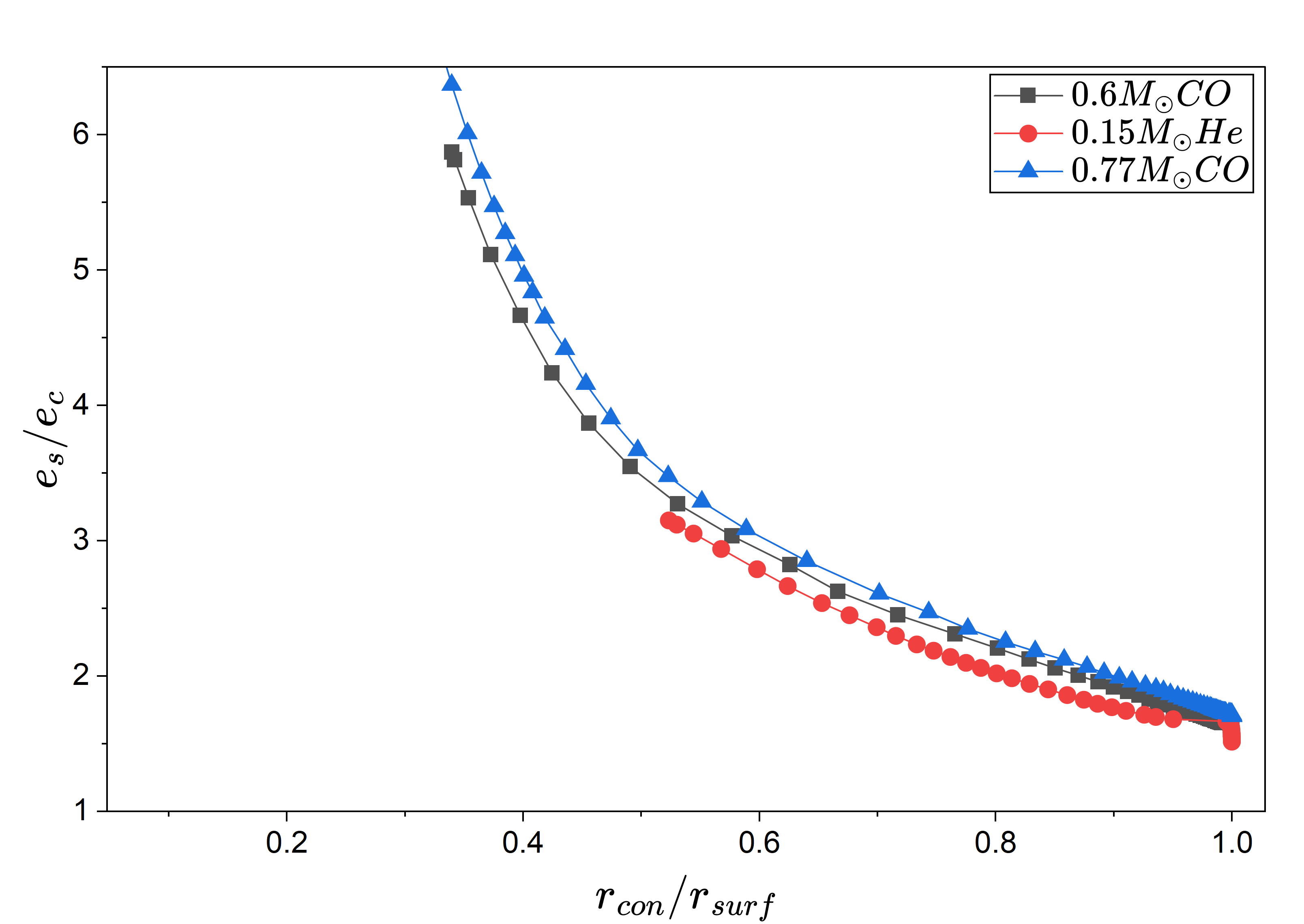}
    \caption{the relationship between the degree of degeneracy and the amplitude of the eccentricity variation}
    \label{5}
\end{figure}

To quantify the extent of the non-degenerate envelope and its direct impact on the interior geometry, we introduce the connection radius $r_{\text{con}}$, defined as the boundary where the Fermi temperature transitions from exceeding to falling below the local temperature. Specifically, the total pressure in the white dwarf can be parameterized as $P = P_{\text{deg}} + P_{\text{th}}$, where $P_{\text{deg}}$ is the degenerate electron pressure and $P_{\text{th}} = \rho k_B T / (\mu m_u)$ represents the ideal gas thermal pressure. The connection radius partitions the star into two distinct regions:
\begin{equation}
P \approx 
\begin{cases} 
P_{\text{deg}},  \text{for } r < r_{\text{con}}\\
\quad (\text{Degeneracy Dominated}) \\
P_{\text{deg}} + P_{\text{th}},\text{for } r > r_{\text{con}}\\
\quad (\text{Thermal Contribution Non-negligible})
\end{cases}
\end{equation}
Within the core region ($r < r_{\text{con}}$), the high degree of electron degeneracy maintains a rigid structure that is relatively insensitive to temperature changes. Conversely, in the outer shell ($r > r_{\text{con}}$), the non-negligible thermal pressure modifies the local polytropic index and alters the structural stiffness. 

As a consequence of the high initial temperature in young white dwarfs, this connection radius $r_{\text{con}}$ retreats significantly toward the stellar center (as illustrated by the low value of $r_{\text{con}}/r_{\text{surf}}$ in Fig.~\ref{5}). This inward migration effectively thickens the outer non-degenerate envelope. Because this thick, thermally-supported shell responds more dramatically to rotational forces than the degenerate core, it breaks the spatial uniformity of the eccentricity distribution, ultimately triggering the non-self-similar geometric deformation of isodensity surfaces.

Moreover, we find that even though the actual temperatures differ significantly among the three white dwarfs (0.15$M_\odot$ He-core red line,  0.6$M_\odot$ CO-core black line, and 0.77$M_\odot$ CO-core blue line), the amplitude of the eccentricity variation remains nearly the same for a given degree of degeneracy. This further confirms that the competition between the local temperature and the Fermi temperature plays a dominant role in determining the amplitude of the eccentricity variation.

\subsection{Results and Analysis}
\label{sec:results}
\begin{figure}[h]
    \centering
    \includegraphics[scale=0.5]{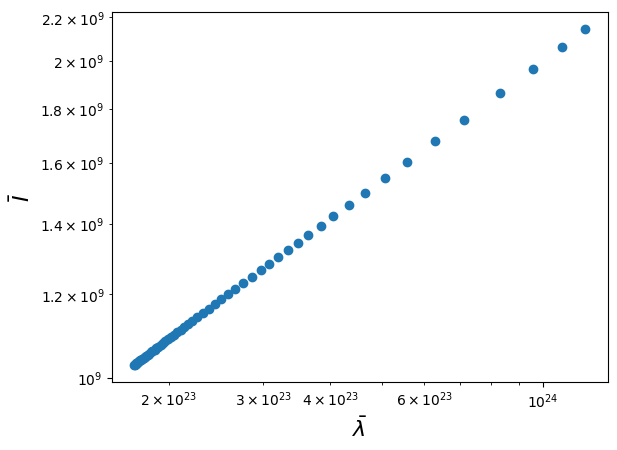}
    \caption{the I-Love relation of the $0.15M_\odot$ He-core white dwarf}
    \label{3}
\end{figure}

\begin{figure}[h]
    \centering
    \includegraphics[scale=0.5]{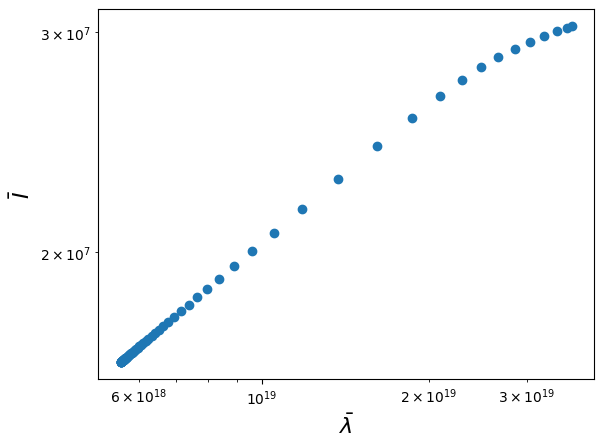}
    \caption{the I-Love relation of the $0.6M_\odot$ CO-core white dwarf}
    \label{4}
\end{figure}

Fig \ref{3} and \ref{4} show the I-Love curves for helium-core and carbon-oxygen core white dwarfs at different temperatures, respectively. In these figures, the direction of decreasing $\bar{I}$ corresponds to the cooling process of the white dwarf. It can be seen that as the white dwarf cools ($\bar{I}$ decreases), $\bar{\lambda}$ also decreases monotonically. At low temperatures, $\bar{I}$ and $\bar{\lambda}$ exhibit an approximately linear relationship. However, in Fig \ref{4}, for the carbon-oxygen core white dwarf at a temperature close to $10^8\,\mathrm{K}$, the I-Love relation clearly deviates from linearity.

To further investigate the effect of temperature variations on the I-Love relation of white dwarfs, we use the Chandrasekhar zero-temperature EOS as a reference. This equation assumes completely degenerate electrons and is applicable to cold white dwarfs. Table \ref{tab:zero_temp_ilove} lists the dimensionless moment of inertia $\bar{I}$ and dimensionless tidal Love number $\bar{\lambda}$ calculated using this EOS (K. Boshkayev et al. 2018) \cite{boshkayev2018non}.

\begin{table}[htbp]                 

  \centering                        
  
  \caption{For the zero-temperature white dwarf model, the I-Love data are from Ref. \cite{boshkayev2018non}. Here, $\rho$ is the central density, $\bar{I}$ is the dimensionless moment of inertia of the static configuration, and $\bar{\lambda}$ is the dimensionless tidal Love number of the static configuration.}         
\label{tab:zero_temp_ilove}
\begin{tabular}{c c c}
\toprule
$\rho\ (\mathrm{g\,cm}^{-3})$ & $\bar{I}$ & $\bar{\lambda}$ \\
\midrule
$10^{2}$   & $1.0 \times 10^{13}$ & $1.7 \times 10^{33}$ \\
$10^{3}$   & $4.8 \times 10^{11}$ & $8.2 \times 10^{29}$ \\
$10^{4}$   & $2.3 \times 10^{10}$ & $4.0 \times 10^{26}$ \\
$10^{5}$   & $1.1 \times 10^{9}$  & $2.2 \times 10^{23}$ \\
$10^{6}$   & $6.7 \times 10^{7}$  & $1.9 \times 10^{20}$ \\
$10^{7}$   & $6.2 \times 10^{6}$  & $4.9 \times 10^{17}$ \\
$10^{8}$   & $9.3 \times 10^{5}$  & $4.5 \times 10^{15}$ \\
$10^{9}$   & $1.8 \times 10^{5}$  & $8.2 \times 10^{13}$ \\
$10^{10}$  & $3.9 \times 10^{4}$  & $1.8 \times 10^{12}$ \\
$10^{11}$  & $8.5 \times 10^{3}$  & $4.0 \times 10^{10}$ \\
\bottomrule
\end{tabular}
\end{table}

\begin{figure}[h]
    \centering
    \includegraphics[scale=0.5]{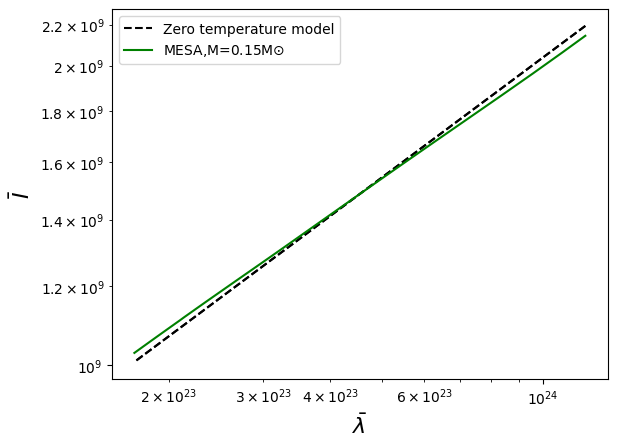}
    \caption{Comparison of I-Love curves between the zero-temperature model and a $0.15\,M_{\odot}$ helium-core white dwarf.}
    \label{fig6}
\end{figure}

\begin{figure}[h]
    \centering
    \includegraphics[scale=0.5]{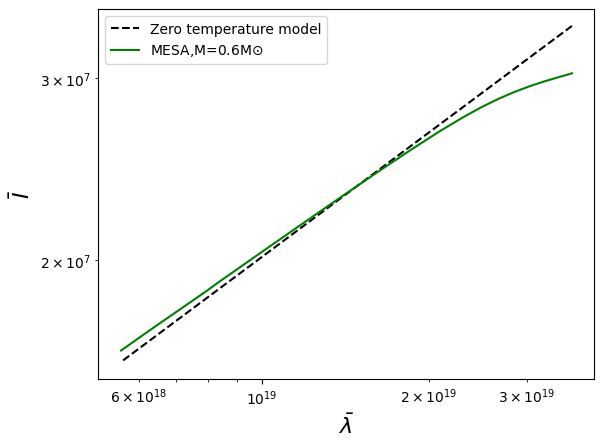}
    \caption{Comparison of I-Love curves between the zero-temperature model and a $0.6\,M_{\odot}$ carbon-oxygen core white dwarf.}
    \label{fig7}
\end{figure}

In Figs. \ref{fig6} and \ref{fig7}, we show the comparison of the I-Love relations between the zero-temperature model and the $0.15\,M_{\odot}$ helium-core white dwarf as well as the $0.6\,M_{\odot}$ carbon-oxygen core white dwarf, respectively. The black dashed curves correspond to the data in Table \ref{tab:zero_temp_ilove} (the zero-temperature model), while the green solid curves are obtained from MESA simulation data.

For He-core white dwarf (Fig. \ref{fig6}), the MESA models at all temperatures agree well with the zero-temperature curve, with relative errors within $2\%$. As the He-core white dwarf has a bigger degree of degeneracy in Fig. \ref{5}, the thermal effect is weak, which leads to a small amplitude of the eccentricity variation ($e_s/e_c<3.2$). So the deviation from the I-Love relation of the zero-temperature model is small.

For the 0.6$M_\odot$ carbon-oxygen white dwarf (Fig. \ref{fig7}), in the high-temperature stage ($T_c$= $10^8$ K, corresponding to an age of $4.28 \times 10^8$ yr), the maximum deviation of the I-Love curve from the zero-temperature model in the MESA model exceeds 10$\%$. As the carbon-oxygen core white dwarf rapidly cools down ($T_c$ drops below $10^7$ K), the curve gradually approaches the zero-temperature model. Consequently, as shown in Figs. \ref{5} and \ref{fig7}, the thermal effects on the structure of the carbon-oxygen core white dwarf at high temperatures are significant, with the variation in the eccentricity of isodensity surfaces reaching a factor of 5.8 at the high-temperature end ($e_s/e_c$= 5.8), thereby breaking the universality of the I-Love relation.

\begin{figure}[h]
    \centering
    \includegraphics[scale=0.3]{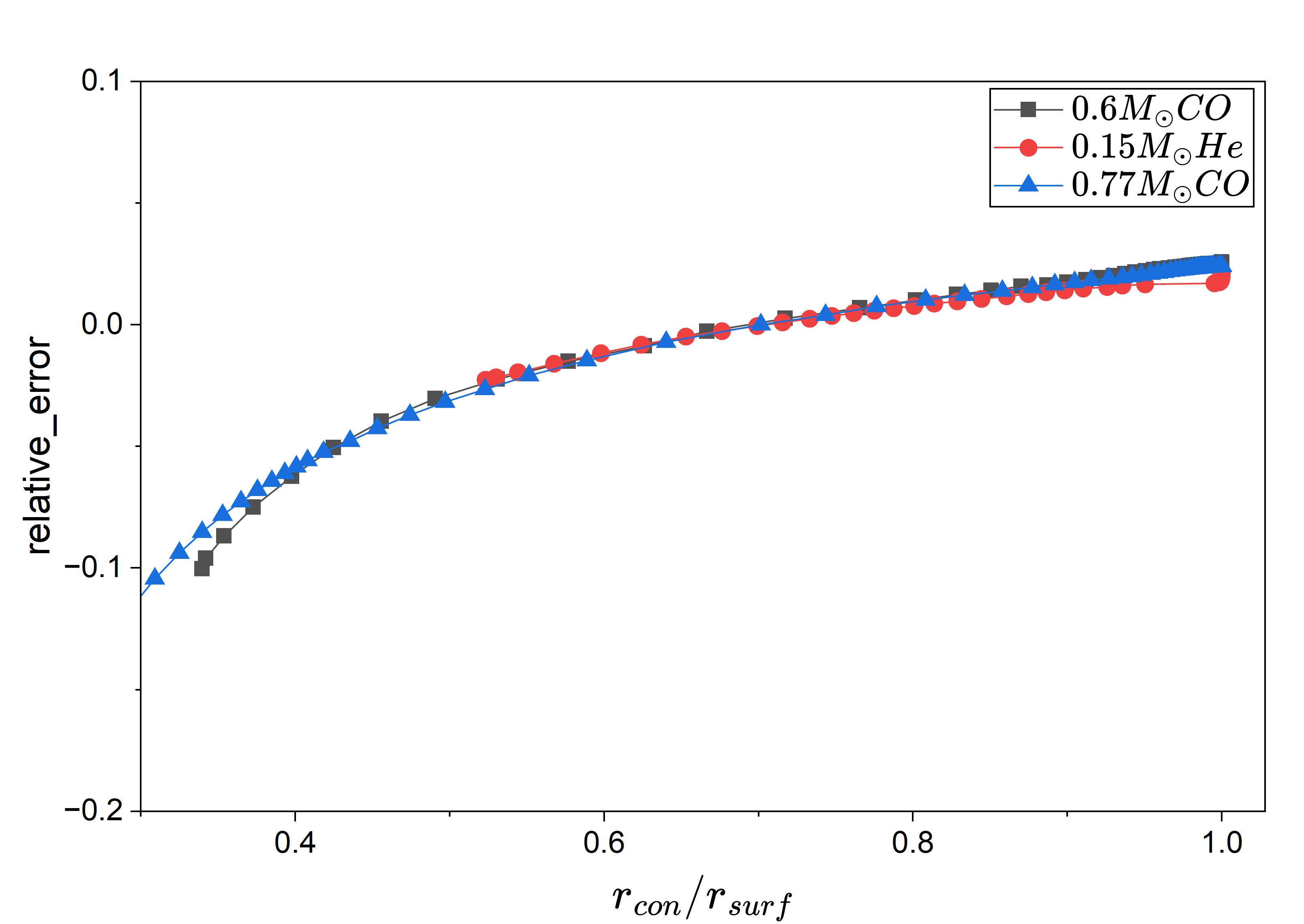}
    \caption{the relationship between the degree of degeneracy and relative error of I-Love relation}
    \label{fig8}
\end{figure}

Furthermore, Fig. \ref{fig8} demonstrates how the relative error in the I-Love relation (between MESA models and the zero-temperature baseline) depends on the degree of degeneracy. Specifically, the degeneracy, quantified by the fractional radius interior to which the Fermi temperature surpasses the local temperature, is found to correspond nearly one-to-one with the relative error. We also observe that the relative error grows as the degeneracy declines.

Combining Figs. \ref{5},\ref{fig6},\ref{fig7} and \ref{fig8}, one can observe that for real white dwarfs, as they gradually cool down, the degree of degeneracy increases, and the variation amplitude of the eccentricity from the interior to the exterior decreases continuously, so the I-Love curve gradually approaches that of the zero-temperature model. This result further verifies the core viewpoint proposed in \cite{yagi2014love}, and agrees well with the simple test we make in Section.~\ref{4.3.3}, namely, the smaller the variation of the eccentricity of isodensity surfaces, the better the universality of the I-Love-Q relations. Furthermore, we demonstrate that the effect of temperature on the I-Love relations essentially stems from the competition between local temperature and Fermi temperature. This competition influences the eccentricity of the isodensity surfaces, ultimately altering the I-Love relations.

\section{Conclusion and Outlook}
Based on hydrostatic equilibrium equations in the Newtonian approximation, this paper outlines static stellar solutions, two typical EOS (polytropic and the Chandrasekhar zero-temperature models), and the computational frameworks for the moment of inertia, tidal Love number, and rotating quadrupole moment. A preliminary test in Section.\ref{4.3.3} establishes that $I\text{--}\text{Love}\text{--}Q$ deviations are dictated by the loss of self-similarity in internal isodensity surfaces.

Using the MESA stellar evolution code, we model $0.15 M_{\odot}$ helium-core and $0.6 M_{\odot}$ carbon-oxygen-core white dwarfs across evolutionary tracks from early hot phases to late cooled states. By solving the Clairaut–Radau equation, we map the radial eccentricity profiles of internal layers, compute the dimensionless moment of inertia $\bar{I}$ and tidal Love number $\bar{\lambda}$, and isolate the role of thermal effects on white dwarf $I\text{--}\text{Love}$ relations. For carbon-oxygen models at high central temperatures ($T_c \sim 10^{8}~\mathrm{K}$), the eccentricity variation amplitude $e_{s}/e_{c}$  reaches $5.8$, signaling pronounced outer ellipsoidal deformation that compromises geometric self-similarity and drives $I\text{--}\text{Love}$ deviations past $10\%$ relative to the zero-temperature baseline. As cooling reduces $T_c$ below $10^{7}~\mathrm{K}$, $e_{s}/e_{c}$ relaxes to $1.6$, smoothly restoring the curves to zero-temperature profiles. Conversely, helium-core white dwarfs experience a lower maximum central temperature ($T_c \sim 2.5\times10^{7}~\mathrm{K}$); their highly degenerate interiors suppress thermal pressure contributions, restricting eccentricity variations ($e_{s}/e_{c} < 3.2$) and keeping $I\text{--}\text{Love}$ residuals under $2\%$.

Our findings substantiate the structural foundation proposed in \cite{yagi2014love} regarding isodensity self-similarity and universality. We show that thermal violations of these relations are not born from an isolated thermodynamic process, but rather a chain reaction: elevated temperatures lower electron degeneracy, distort the density profile, and amplify the radial variation of isodensity eccentricity. Ultimately, across neutron stars, white dwarfs, and main-sequence stars alike, the true metric for $I\text{--}\text{Love}\text{--}Q$ universality is the eccentricity gradient of internal layers rather than the specific EOS. This framework offers fresh numerical validation and physical parameters clarifying the boundaries of these universal relations.

Future trajectories will focus on formalizing the mathematical mapping between degeneracy and eccentricity profiles, alongside extending calculations to more massive carbon–oxygen and neon–oxygen–magnesium white dwarfs.

\section*{Acknowledgments}
The authors would like to thank  Chian-Shu Chen, Alessandro Parisi and Tian-Shun Chen for helpful discussions. K.Z. (Hong Zhang) is supported by a classified fund from Shanghai city.

\bibliographystyle{unsrt} 
\bibliography{WD} 

\end{document}